\documentclass[aps,prl,twocolumn,superscriptaddress,showpacs]{revtex4-1}

\usepackage{graphicx}
\usepackage{dcolumn}
\usepackage{bm}

\usepackage{epstopdf}
\usepackage{hyperref}

\usepackage{colortbl}
\hypersetup{colorlinks=true, citecolor=blue, linkcolor=blue, urlcolor=blue}

\begin{document}


\title{Magnetic phase diagram of CuO}



\author{R. Villarreal}
\email[Email address: ]{renan.villarreal@mun.ca}

\author{G. Quirion}

\author{M.L. Plumer}
\affiliation{Department of Physics and Physical Oceanography, Memorial University, St. John's, Newfoundland, Canada, A1B 3X7}

\author{M. Poirier}
\affiliation{D\'{e}partement de Physique, Univesit\'{e} de Sherbrooke, Sherbrooke,
Qu\'{e}bec J1K 2R1, Canada}

\author{T. Usui}

\author{T. Kimura}
\affiliation{Division of Materials Physics, Osaka University, Toyonaka, Osaka, Japan}


\date{\today}

\begin{abstract}
High resolution ultrasonic velocity measurements have been used to determine the
temperature -- magnetic-field phase diagram of the monoclinic multiferroic CuO. A new
transition at $T_{N3}=230$~K, corresponding to an intermediate state between the
antiferromagnetic non-collinear spiral phase observed below $T_{N2} = 229.3$~K  and the
paramagnetic phase, is revealed. Anomalies associated with a first order transition to
the commensurate collinear phase are also observed at $T_{N1} = 213$~K. For fields with
${\bf B \parallel b}$, a spin-flop transition is detected between 11~T - 13~T at lower
temperatures. Moreover, our analysis using a  Landau-type free energy clearly reveals the
necessity for an incommensurate collinear phase between the spiral and the paramagnetic
phase. This model is also relevant to the phase diagrams of other monoclinic multiferroic
systems.
\end{abstract}


\maketitle


Multiferroic phenomena have been a subject of intense interest in recent decades arising
from opportunities to explore new fundamental physics as well as possible technological
applications \cite{ain92,Fiebig05,kimura08}. Coupling between different ferroic orders
has been proven to be driven by several different types of mechanisms. In particular,
multiferroics with a spiral spin-order-induced ferroelectricity have revealed high
spontaneous polarization and strong magnetoelectric coupling \cite{kimura03,wang09}.
Cupric oxide (CuO), the subject of this letter, was characterized as a magnetoelectric
multiferroic four years ago when it was shown that its ferroelectric order is induced by
the onset of a spiral antiferromagnetic (AFM) order at an unusually high temperature of
230~K \cite{kimura08}.
Thus far, two AFM states have been reported, a low temperature ($T_{N1}\sim 213$ K) AF1
commensurate collinear state with the magnetic moments along the monoclinic ${\bf b}$
axis and an AF2 incommensurate spiral state with half of the magnetic moments in the
${ac}$ plane ($T_{N2}\sim 230$ K) \cite{kimura08, forsyth88, Babkevich12}. However, the
authors of the neutron diffraction measurements \cite{forsyth88} questioned the
possibility of having a direct condensation from a paramagnetic (PM) phase to a spiral
magnetic phase. Despite this remark, a recent Landau theory \cite{toledano11}, as well as
several Monte-Carlo simulations \cite{giovannetti11, Jin12}, appear to support this
sequence of magnetic orderings.

Encouraged by recent experiments on other multiferroic systems using ultrasonic
measurements \cite{quirion09}, we measured the temperature and field dependence of the
velocity of transverse modes in order to determine the magnetic phase diagram of CuO. A
new transition is detected at $T_{N3}=230$~K just above the AF2 spiral phase observed at
$T_{N2}=229.3$~K, while the first order transition is observed at $T_{N1}=213$~K.
Furthermore, dielectric constant measurements confirm that only the spiral phase (between
$T_{N1}$ and $T_{N2}$) supports a spontaneous electric polarization. In addition, we
report on a spin-flop transition in the low temperature AF1 collinear phase when ${\bf
B\parallel b}$. Thus, based on these findings, a new magnetic-field {\it vs} temperature
phase diagram is proposed for CuO.

In order to elucidate the possible nature of the AFM states observed in CuO, a non-local
Landau-type free energy is also developed for CuO and similar monoclinic multiferroics.
This approach has been very successful in explaining the magnetic phase diagrams of other
multiferroic systems \cite{plumer08,condran10,quirion11}. In contrast with the
conclusions of Refs.~\cite{toledano11, giovannetti11, Jin12}, our analysis based on
rigorous symmetry arguments indicates that there must be a collinear intermediate phase
(AF3) between the paramagnetic and spiral AF2 states. Such a phase has been shown, both
theoretically and experimentally, to occur in other geometrically frustrated
antiferromagnets where symmetry allows for uniaxial anisotropy at second
order~\cite{plumer88b, quirion11}. Finally, we compare the model predictions to the B-T
phase diagram of CuO obtained using ultrasonic velocity data. Similarities with other
multiferroic systems such as MnWO$_4$, AMSi$_2$O$_6$, RMnO$_3$, RMn$_2$O$_5$, and
Ni$_3$V$_2$O$_8$ are also noted.

For the purpose of this study, a CuO sample was grown using a floating zone technique as
described in Ref.~\cite{kimura08}. A single crystal was cut with faces perpendicular to
the monoclinic axes ${\bf a^*}$, ${\bf b^*=b}$, and ${\bf c^*}$ ($4\times4\times3$
mm$^3$). The sample was then polished to obtain parallel faces. For velocity
measurements, plane acoustic waves were generated using 30 MHz LiNbO$_3$ piezoelectric
transducers bonded to opposite faces. Using an ultrasonic interferometer, which measures
the phase shift and the amplitude of the first elastic transmitted pulse, high-resolution
relative velocity variations ($\Delta V/V \sim 1$ ppm) were achieved. Experimental data
presented here were all obtained using the velocity of transverse waves V$_{a^*}$[$c^*$]
propagating along the ${\bf a^*}$ axis and polarized along ${\bf c^*}$, with the magnetic
field applied along the easy magnetic axis of CuO (${\bf b}$ axis). Simultaneous
capacitance measurements were carried out using an AH 2550A Ultra Precision 1kHz
Capacitance Bridge to identify which of these phases are ferroelectric. For that purpose,
electrodes were mounted on faces perpendicular to the ${\bf b}$ axis in order to
determine the dielectric constant $\epsilon{_b}$.
\begin{figure}[t]
\includegraphics[width=1\columnwidth]{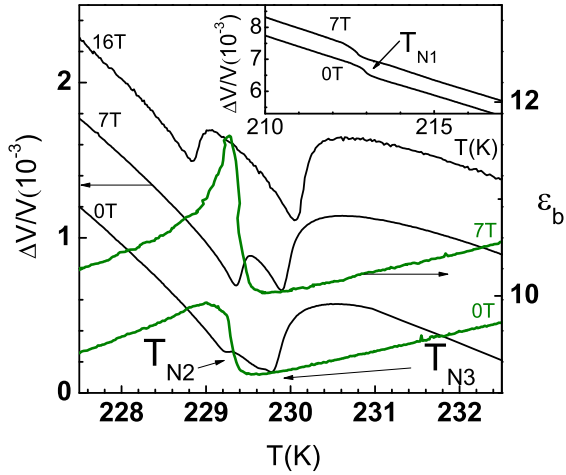}
\caption{\label{data} Temperature dependence of the dielectric constant $\epsilon{_b}$
(in green) and the relative velocity variations of transverse mode V$_{a^*}$[$c^*$]
measured at different fields with ${\bf B\parallel b}$.}
\end{figure}

\begin{figure}[b]
\includegraphics[width=1\columnwidth]{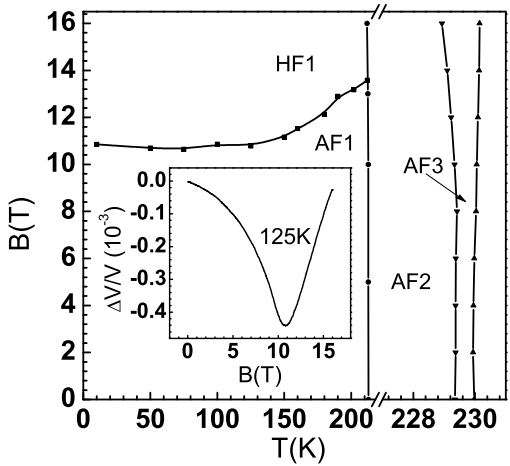}
\caption{\label{phasediagram} Magnetic phase diagram of CuO for {\bf B}$\parallel$ {\bf
b}. Inset shows the relative velocity variation of V$_{a^*}$[$c^*$] as a function of the
field for T = 125~K.}
\end{figure}
Fig.~\ref{data} shows the temperature dependence of the relative sound velocity
variations ($\Delta V/V$) for ${\bf B\parallel b}$. At zero field, the anomaly observed
at $T_{N1} = 213$~K (see inset of Fig.~\ref{data}) coincides very well with the onset of
a commensurate collinear antiferromagnetic state. Our high resolution velocity
measurements also reveal \emph{two} anomalies at $T_{N2} = 229.3$~K and $T_{N3} =
230.0$~K near the stabilization of a spiral order previously determined by neutron
diffraction and susceptibility measurements \cite{forsyth88,kimura08}, which were thought
to occur at a single transition.  At higher fields, the amplitude of the step like
variation observed at 229.3~K, as well as the temperature difference between $T_{N2}$ and
$T_{N3}$ increases, confirming the existence of a new intermediate magnetic order AF3.
This finding is supported by dielectric measurements also shown in Fig.~\ref{data}.
Notice that, as the stability range of the intermediate phase is small ($\Delta T \sim
0.7$~K), velocity and dielectric data have been collected simultaneously to avoid any
ambiguity regarding the actual critical temperatures. Thus, as shown in Fig.~\ref{data}
(for B = 0 and 7~T), the anomaly observed on the dielectric constant $\epsilon_b$
coincides very well with $T_{N2}$ determined using velocity data, while no variation is
noticeable at $T_{N3}$. These results also indicate that the new phase AF3 is not
ferroelectric, while magnetoelectric coupling exists for the AF2 phase.

We present in Fig.~\ref{phasediagram} the magnetic phase diagram of CuO determined up to
16~T using ultrasonic velocity measurements for \textbf{B} $\parallel$  \textbf{b}. The
inset of Fig.~\ref{phasediagram} shows the field dependence of the velocity which
displays a minimum around 11~T for $T=125$~K. As the magnetic moments are known to be
parallel to the field in the AF1 commensurate collinear state \cite{kimura08, forsyth88},
we attribute this anomaly to a spin-flop transition \cite{quirion06}. In summary, while
the critical temperatures $T_{N1}$, $T_{N2}$, and $T_{N3}$ are weakly field dependent,
the spin-flop critical field $H_{SF}$ increases with temperature. At 10 K, $H_{SF} =
11$~T and increases slowly up to 13.5~T at $T_{N1}$, in good agreement with magnetic
susceptibility measurements performed on powder samples \cite{kondo88}.

\begin{figure}[b]
\includegraphics[width=1\columnwidth]{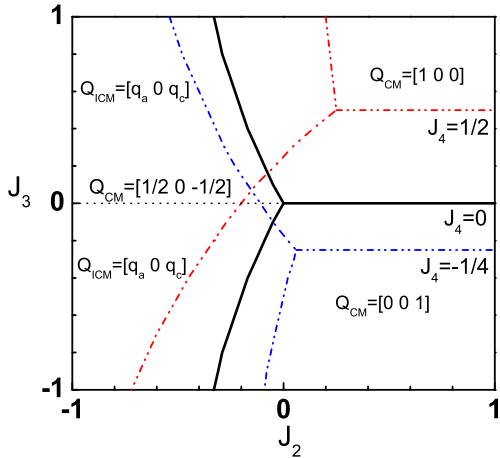}
\caption{$J_3-J_2$ phase diagram for different values of $J_4$ with $J_1 = 1$. One
incommensurate phase with $\textbf{Q}_{ICM} = [q_a,0,q_c]$ (left side) and two
commensurate phases (${\bf Q}_{CM}=[100]$ and  ${\bf Q}_{CM}=[001]$) are obtained.}
\label{js}
\end{figure}

\begin{figure}
\includegraphics[width=1\columnwidth]{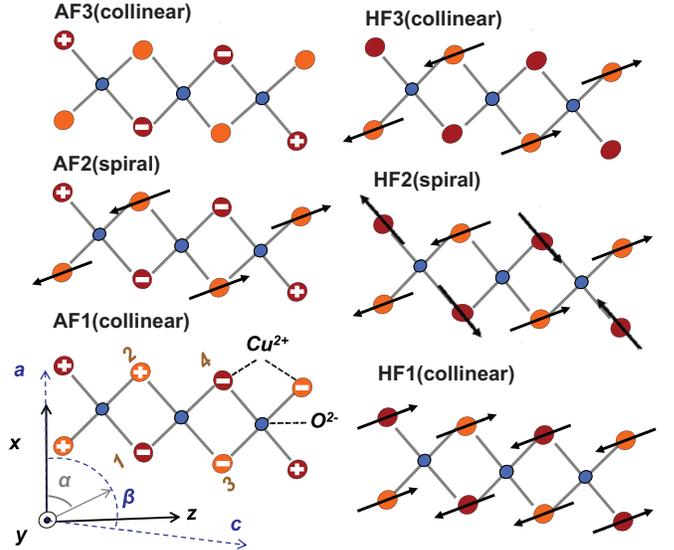}
\caption{Spin configurations in a magnetic cell of 8 ions (red and orange circles). Red
circles represent magnetic ions at b = 1/2. The +/- symbols represent spins in/out of the
page. When no direction is specified (as in AF3 and HF3), spins on these sites are not
ordered.} \label{fig:configuration}
\end{figure}

Since no neutron scattering data exists for the HF1 and AF3 states, we develop a
Landau-type model in order to elucidate the nature of these new magnetic orders
\cite{plumer88b, quirion11}. The integral form of the free energy is expanded in powers
of the nonlocal spin density {\bf s}({\bf r}) defined in terms of a uniform field-induced
magnetization ${\bf m}$ and a spin polarization vector {\bf S} modulated by a single wave
vector ${\bf Q}$ describing the long-range magnetic order (Eq.~(6) of
Ref.~\cite{quirion11}). Within the present model, the value of ${\bf Q}$ can be
determined by simply considering the isotropic quadratic contribution
\begin{eqnarray}
F_{2I} = \frac{1}{2V^2}\int d{\bf r_1} d{\bf r_2} A({\bf r_1-r_2}) {\bf s(r_1)}\cdot {\bf
s(r_2)},
\end{eqnarray}
which leads to $F_{2I} = \frac{1}{2} \tilde{A} m^2 + A_Q S^2$ where $A_Q = a T + J_Q$,
with $J_Q$ being the Fourier transform of the exchange integral $J({\bf R})$. Considering
the C-type monoclinic cell with four Cu$^{2+}$ magnetic ions, we obtain
\begin{eqnarray}\label{eq:Q}
J({\bf Q})&=&2\left[J_1f_1({\bf Q}) + J_2f_2({\bf Q}) + J_3f_3({\bf Q}) + J_4f_4({\bf Q}) \right] \nonumber \\
f_1({\bf Q})&=&\cos{(\pi q_a - \pi q_c)} \nonumber \\
f_2({\bf Q})&=&\cos{(\pi q_a + \pi q_c)}  \\
f_3({\bf Q})&=&\cos{(\pi q_a - \pi q_b)} + \cos{(\pi q_a + \pi q_b)} \nonumber \\
f_4({\bf Q})&=&\cos{(\pi q_b - \pi q_c)} + \cos{(\pi q_b + \pi q_c)}, \nonumber
\end{eqnarray}
where $J_1$ and $J_2$ represent the nearest-neighbors (NN) exchange interactions along
the AFM-chain (sites 2-3) and the coupling between chains (sites 1-4) on the same plane
normal to {\bf b}, respectively, and $J_3$ and $J_4$ represent the exchange interactions
along \textbf{a} (sites 1-2) and \textbf{c} (sites 1-3) between ions on different planes
(see Fig.~\ref{fig:configuration}). The value of ${\bf Q}$ is then obtained by finding
the extrema of $J_Q$ (Eq.~\ref{eq:Q}) as a function of the exchange interactions. Results
of our numerical algorithm are summarized in the $J_2-J_3$ phase diagram shown in
Fig.~\ref{js} for AFM chains ($J_1=1$). For different $J_4$ values, we obtain three
phases: an incommensurate phase with $\textbf{Q}_{ICM} = [q_a,0,q_c]$ (left side) and two
commensurate phases (top and bottom right side). Depending on the sign of $J_3$ relative
to $J_4$, the commensurate wave vector is either $\textbf{Q}_{CM} = [100]$ or
$\textbf{Q}_{CM} = [001]$.
More interestingly, with $J_3 = J_4 = 0$ we obtain the expected commensurate wave vector
${\bf Q}_{CM}=[\frac{1}{2}\ 0\ -\frac{1}{2}]$ for $J_2 \leq 0$ (dash line in
Fig.~\ref{js}).  Moreover, an ICM state with a modulation vector comparable to that of
the experimental value ${\bf Q}_{ICM}=[0.506\ 0\ -0.483]$ is stabilized whenever $J_3$
and/or $J_4$ are non-zero but small relative to $J_1$ (for example, $J_2/J_1=-0.3$,
$J_3/J_1=0.017$, and $J_4/J_1=0$ leading to $J_Q/J_1=-2.6$).  These relative values are
also in good agreement with estimates obtained by density functional theory
\cite{giovannetti11, rocquefelte11, pradipto12} and are consistent with the quasi-1D
magnetic character of CuO.

In addition to the usual isotropic second order exchange term, we also consider
anisotropic contributions.  Considering the symmetry of monoclinic crystals ($C2/c$), we
identified three invariants, written in single-ion form as
\begin{eqnarray}
F_{2A} & = & \frac{1}{2V}  \int  \left[ D_y({\bf r}) s_y{\bf (r)} s_y{\bf (r)} +
  D_z({\bf r}) s_z{\bf (r)} s_z{\bf (r)} \right. \nonumber \\
      && ~~~~~~~~ + \left. D_{xz}({\bf r}) s_x{\bf (r)} s_z{\bf (r)} \right] d{\bf r} ~.
\end{eqnarray}
While $D_y$ can be used to set the magnetic easy axis along \textbf{b}, the other terms
are necessary in order to define the direction of the moments in the $ac$ plane.
Furthermore, to account for non-collinear spin configurations, we define ${\bf S}={\bf
S_1} + i~{\bf S_2}$, with
\begin{eqnarray}
{\bf S_1}&=&S\cos{\beta}[\cos{\gamma}~{\bf\hat{y}}+\sin{\gamma}~{\bf\hat{\rho}_2}],\\
{\bf S_2} &=& S \sin{\beta}
[\cos{\theta}~{\bf\hat{\rho}_1}+\sin{\theta}(\cos{\gamma}~{\bf\hat{y}}+\sin{\gamma}~{\bf\hat{\rho}_2})],
\nonumber
\end{eqnarray}
where ${\bf\hat{\rho}_1}$ and ${\bf\hat{\rho}_2}$ are two orthogonal unit vectors normal
to the easy axis, ${\bf \hat{y}\parallel b}$. Thus, the direction of the moments in the
$ac$ plane is accounted for by defining the unit vectors ${\bf \hat{\rho}_1}$ and ${\bf
\hat{\rho}_2}$ relative to the lattice vectors, ${\bf
\hat{\rho}_1}=\cos{\alpha}~{\bf\hat{x}}+\sin{\alpha}~{\bf\hat{z}}$ and ${\bf
\hat{\rho}_2}=-\sin{\alpha}~{\bf\hat{x}}+\cos{\alpha}~{\bf\hat{z}}$. As shown in
Fig.~\ref{fig:configuration}, the parameter $\alpha$ represents the angle between the
$ac$ plane component of {\bf S} relative to the monoclinic axis {\bf a} $\parallel
\hat{x}$. After integration, all second-order contributions for ${\bf m} \parallel {\bf
H}
\parallel \hat{y}$ reduce to
\begin{eqnarray}
F_2^{total} &=& \frac{1}{2}\tilde{A}_0m^2 + A_{\bf Q}S^2 -
\frac{1}{2}D_{y0}m^2-D_{yQ}|S_y|^2\nonumber\\
&-& D_{zQ}|S_z|^2 + D_{xzQ}S_xS_z - {\bf H\cdot m}. \label{f2}
\end{eqnarray}
Adopting the same approach for the fourth-order isotropic term, we obtain
\begin{eqnarray}
F_{4I} &=& B_1S^4 +\frac{1}{2}B_2|{\bf S\cdot S}|^2 + \frac{1}{4}B_3m^4+2B_4|{\bf m\cdot S}|^2 \nonumber \\
&+& B_5m^2S^2 + \frac{1}{4}B_U[({\bf S\cdot S})^2 + c.c.] \Delta_{4{\bf Q,G}}~.
\label{eq:F4}
\end{eqnarray}
Note the umklapp term $\Delta_{4{\bf Q,G}}$, arising directly from the lattice
periodicity \cite{plumer08}. This term is crucial in order to account for the first order
phase transition observed at $T_{N1}$ in CuO where a commensurate collinear state is
stabilized.

The free energy, $F = F_{2I}+F_{2A} + F_{4I}$, with $A_Q = a (T -T_Q)$ and
$\tilde{A}_0-D_{y0}=a(T-T_0)$, is then numerically minimized. As in
Ref.~\cite{plumer88b}, most coefficients are set using analytical solutions associated
with phase boundaries of second order transition. For example, setting $T_Q = 1.18$,
$D_{yQ} = 0.02$, $B_1=0.103$, and $B_2=0.011$, reasonable values for the critical
temperatures at zero field ($T_{N3} = 1.2$ and $T_{N2} = 1.12$). We also set
$D_{zQ}=0.01$ as we must have $D_{zQ} < D_{yQ}$, while the direction of the spins in the
$ac$ plane ($\alpha_{exp} \sim 70^\circ$) \cite{ain92} is used to determine the ratio
$D_{xzQ}/D_{zQ} = - 0.42$. The last coefficients are determined using the temperature of
the multicritical point (where $T_{N2}$ and $T_{N3}$ boundaries meet) and the maximum
field at $T = 0$ K. From this exercise, we find $B_3=0.063$ and $B_4=0.013$ while $B_5 =
0.1$ was set arbitrarily. Finally, $B_U=0.035$ is used to obtained $T_{N1} = 0.77$.
\begin{figure}[b]
\includegraphics[width=1\columnwidth]{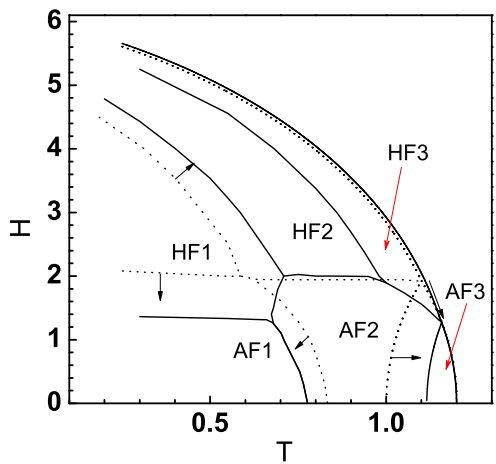}
\caption{Magnetic field - temperature phase diagram of CuO for ${\bf H\parallel b}$
derived from the Landau free energy. Dotted lines represent prediction with only one
anisotropic term included, $D_{yQ}$. The solid line is for the case where all anisotropic
terms considered.} \label{thphd}
\end{figure}

Fig.~\ref{thphd} shows the magnetic phase diagram obtained from minimization of the free
energy. For comparison, we also present results obtained without the anisotropic terms
$D_{zQ}$ and $D_{xzQ}$ (dotted lines). Depending on the scenario considered, we obtain 5
or 6 magnetic phases illustrated in Fig.~\ref{fig:configuration}, described by the order
parameters listed in Table~\ref{table:op}. At zero field, both models (with and without
$D_z$ and $D_{xz}$) predict the same phase sequence, consistent with our experimental
observations shown in Fig.~\ref{phasediagram}. At low temperatures, a collinear phase AF1
with the moments along \textbf{b} is predicted (see Fig.~\ref{fig:configuration}) while
the AF2 phase corresponds to a spiral configuration in agreement with neutron scattering
data \cite{forsyth88}. According to our numerical calculation, the new intermediate phase
AF3 is associated with a collinear phase where only half of the moments order with
\textbf{S} $\parallel$ \textbf{b}. As the field is applied, two spin-flop transitions
(AF1$\rightarrow$ HF1 and AF2 $\rightarrow$ HF2) are found. The comparison of both phase
diagrams indicates that the role of the anisotropic terms $D_{zQ}$ and $D_{xzQ}$ is to
reduce the critical field of the AF1$\rightarrow$ HF1 transition, decrease the stability
range of the intermediate phase AF3, and lead to a new magnetic order HF3 in which half
the moments align into the $ac$ plane. These findings could account for the fact that no
spin-flop phase transition has been observed experimentally up to 16~T for the spiral
phase AF2.
\begin{table}[t]
\caption{Order parameters.} \label{table:op}
\begin{ruledtabular}
\begin{tabular}{lllll}
state & $\beta$ & $\theta$ & $\gamma$ & $\alpha$\\
\hline
AF1 & $\pi/ 4 $ & $\pi/2$ & 0 & -\\
AF2 & $\beta$ & 0 & 0 & $70^\circ$\\
AF3 & - & - & 0 & -\\
HF1 & $\pi/4 $ & $\pi/2$ & $\pi/2$ & $160^\circ$\\
HF2 & $\beta$ & 0 & $\pi/2$ & $70^\circ$\\
HF3 & $\pi/2 $ & 0 & - & $70^\circ$
\end{tabular}
\end{ruledtabular}
\end{table}

Our principal conclusions are that a new collinear phase (AF3) has been detected by high
resolution ultrasonic velocity measurements which occurs between the paramagnetic and the
previously identified spiral phase.  The magnetic-field vs temperature phase diagram for
\textbf{B} $\parallel$ \textbf{b} has also been determined, revealing the existence of a
new spin-flop phase (HF1). Complementary dielectric measurements also confirm that
magnetoelectric effects only exist in the non-collinear phase. Verification that the new
AF3 phase must exist is achieved by a Landau-type model based on rigorous symmetry
arguments. Furthermore, the occurrence of such a collinear state, just above a
non-collinear state, is confirmed in well studied frustrated RMnO$_3$ and RMn$_2$O$_5$
systems \cite{oflynn11,kimura05,noda08}, and the kagom\'{e} compound Ni$_3$V$_2$O$_8$
\cite{lawes05}. Finally, the proposed model accounts for the experimental phase diagram
of CuO determined in this work and is potentially useful for the description of other
monoclinic multiferroic systems, in particular MnWO$_4$ \cite{felea11} and AMSi$_2$O$_6$
\cite{jodlauk07}.

\section{Acknowledgments}
This work was supported by the Natural Science and Engineering Research Council of Canada
(NSERC).


\end{document}